\documentclass[useASM, referee]{mn2e}
\input{psfig.sty}
\usepackage{graphicx}
\begin{document}
\title{Characteristics of the curvature effect of uniform jets or
spherical fireballs }
\date{2005 February 20}
\pubyear{????} \volume{????} \pagerange{2} \onecolumn
\author[Lu and Qin]
       {R.-J.\ Lu$^{1,2,3,4}$, Y.-P.\ Qin$^{1,2,4}$ \\
$^1$National Astronomical Observatories/Yunnan Observatory,
Chinese Academy of Sciences,\\P. O. Box 110, Kunming 650011, China\\
$^2$Physics Department, Guangxi University, Nanning, Guangxi 530004, P.R. China\\
$^3$The Graduate School of the Chinese Academy of Sciences; luruijing@126.com\\
$^4$E-mail:luruijing@126.com, ypqin@public.km.yn.cn
 }
\pagestyle{plain}
\date{Accepted ????.
      Received ????;
      in original form 2005 August 08}
\pagerange{\pageref{firstpage}--\pageref{lastpage}} \pubyear{2004}
\maketitle \label{firstpage}
\begin{abstract}
Qin et al. (2004) have derived a formula of count rates based on a
model of highly symmetric expanding fireballs, where the Doppler
effect is the key factor to be concerned. In this paper, we employ
the formula to both the Qin model and the uniform jet model to
study how the rising timescale, $\Delta\tau_{\theta,r}$, and the
decay timescale, $\Delta\tau_{\theta,d}$, of a local pulse affect
the light curve of GRBs. Our analysis shows that they do make
contributions to both rising and decay portions of the light curve
of GRBs. Associated with a local pulse with both rising and decay
portions, the light curve of GRBs in the rising portion is
expected to undergo a concave phase and then a convex one, which
we call a ``concave-to-convex" character, whereas that in the
decay portion is expected to evolve an opposite process which is
called a ``convex-to-concave" character, regardless of being in
Qin model or in the uniform jet model. These characteristics are
independent of local pulse shapes and their rest frame radiation
forms. The diversity of light curves are caused by different forms
of local pulses and different ratios of $\Delta\tau_{\theta,r}$ to
$\Delta\tau_{\theta,d}$. We study a sample of 86 GRB pulses
detected by the BATSE instrument on board the Compton Gamma Ray
Observatory and find that the ``concave-to-convex" (expected in
the rising portion of the light curve) and ``convex-to-concave"
(expected in the decay portion) characters, which result from the
curvature effect of spherical surface (or the Doppler effect), do
exist in the light curve of some GRBs.
\end{abstract}

\section{Introduction}
Light curves of gamma-ray bursts (GRBs) vary drastically in
morphology.  only less than 10 percent of the light curve profiles
detected by the BATSE instrument on board the Compton Gamma Ray
Observatory (CGRO) can be categorized as being similar in the
overall shape. A pulse in the gamma-ray light curve is generally
considered to result from an individual shock episode.
Superposition of many such pulses is believed to create the
observed diversity and complexity of light curves, which could not
be describable by a simple formula. Therefore, the temporal
characteristics of these pulses hold the key to the understanding
of the prompt radiation of GRBs.

Some well-separated pulses are seen to comprise a fast rise and an
exponential decay (FRED) (Fishman et al. 1994). Many attempts have
been made to decompose the complex light curves into pulses, which
can be described by several flexible functions based on empirical
or semi-empirical relations, and quantify their characteristics
(e.g., Norris et al. 1996; Femimore et al. 1996; Lee et al. 2000a,
2000b; Ryde \& Svensson 2000, 2002; Ryde \& Petrosian 2002;
Borgonovo \& Ryde 2001; Kocevski et al. 2003, hereafter Paper I).
E.g., as derived in detail in Paper I, a FRED pulse could be well
described by equation (22) or (28), and using this model, they
revealed some features of the profiles of individual GRB pulses.

However what physics causes such a observed profiles? According to
Ryde \& Petrosian (2002), the simplest scenario accounting for the
observed GRB pulses is to assume an impulsive heating of the
leptons and a subsequent cooling and emission. In this scenario,
the rising phase of the pulse, which is referred to as the dynamic
time, arises from the energizing of the shell, while the decay
phase reflects the cooling and its timescale. However, in general,
the cooling time for the relevant parameters is too short to
explain the pulse durations and the resulting cooling spectra are
not consistent with observation (Ghisellini et al. 2000). As shown
by Ryde \& Petrosian (2002), this problem could be solved when the
curvature effect of the expanding fireball surface is taken into
account.

However, there is no consensus on what effects lie behind the
observed pulse shapes. We suspect that the pulse shapes might have
some relations to their counterparts . Motivated by this, we
explore in this paper possible explanations for the pulse shapes
observed based on the curvature effect (or the Doppler effect of
fireballs).

Due to the observed output rate of radiation, the observed
gamma-ray pulses are believed to be produced in a highly
relatively outflow, (e.g., an expanding and collimated fireball),
based on the large energies and the short timescales involved. In
the case of a spherical fireball, the Doppler effect over the
whole fireball must be at work, and the observed FRED structure
probably provide the intrinsical information. Qin (2002) has
derived the flux function based on the model of highly symmetric
expanding fireball, where the Doppler effect of the expanding
fireball surface is the key factor to be concerned, and then with
this formula, Qin (2003) studied how emission and absorbtion lines
are affect by the effect. Recently, Qin et al. (2004) presented
the formula in terms of count rates. Based on their model, many
characteristics of profiles of observed gamma-ray burst pulses
could be explained. They convince us that the exponential decay
phase of the FRED pulse arises from the time delay of different
parts of the surface and its rising portion is produced by the
width of a local pulse. However, some interesting questions, such
as what a role the rising and decay portions of local pulses play
in the observed light curve of gamma-ray bursts, and what are
responsible for the diversity of the FRED pulses, urge us to
investigate in a much detail.

Jets in GRBs were first suggested by Waxman et al. (1998), and
then they were widely evoked to explain its spectacular energy
release (e.g.,Fruchter et al. 1999 ). Subsequent multiwavelength
observations of GRBs have been interpreted as evidence for
explosions with jet-like geometry (Stanek et al. 1999; Harrison et
al. 1999). Recently, two phenomenological models of GRB jets
received widespread attention: a uniform jet model in which the
emissivity is a constant independent of the angle relative to the
jet axis (Frail et al. 2001;Panaitescu \& Kumar 2001; Rossi et al.
2002; Zhang et al. 2002; Bloom et al. 2003; Zhang et al. 2004a,
2004b), and a structure jet model in which the emissivity is a
function of the angle relative to the jet axis (Frail et al. 2001;
Rmairez-Ruiz et al. 2002; Lamb et al. 2005). In the case of a
uniform jet, the formula presented in Qin et al. (2004) could be
applicable as long as the open angle of the jet is taken into
account.

According to the relativistic fireball model, the emission from a
spherically expanding shell and a jet would be rather similar to
each other as long as we are along the jet's axis and the Lorentz
factor $\Gamma$ of the relativistic motion satisfies $\Gamma^{-1}<
\theta_{j}$, because the matter does not have enough time (in its
own rest frame) to expand sideways at such situation (Piran 1995).

The detection of very energetic photons (in the GeV range) in
several gamma-ray bursts (GRBs) and the very short variability
timescale (Fishman et al. 1995), sometimes less than 1 ms,
exhibited by the 100 keV emission of many bursts have led to the
conclusion that they arise from sources that are moving at
extremely relativistic speeds, with very large Lorentz factors
$\Gamma$ that could exceed 100 (Fenimore, Epstein, \&Ho 1993;
Piran 1999). At the same time, From the derived parameter,
$\theta_{j}$, of the sample of 41 source in Friedman et al.
(2004), we find that the range of $\theta_{j}$ is \{0.03107,
0.57072 \} rad, and its average value is 0.15925 rad. The data
show that $\Gamma^{-1}< \theta_{j}$ if $\Gamma>100$. In this
paper, we assume that the condition of $\Gamma^{-1}< \theta_{j}$
is satisfied for all the source concerned.

This paper is organized as follows. In section 2, we present the
characteristics of the light curve of GRBs obtained from our
theoretical analysis based on a uniform jet and a spherical
fireball model. In section 3, we examine the characteristics of
light curves of a group of source detected by the BATSE instrument
on board the Compton Gamma Ray Observatory. Discussion and
conclusions will be presented in the last section.
\section{ Numerical analysis}
As derived in detail in Qin et al. (2004 ), the expected count
rate of the fireball within frequency interval $[\nu_1, \nu_2]$
can be calculated with equation (21), which is as follows:
\begin{equation}
C(\tau )=\frac{2\pi R_c^3\int_{\widetilde{\tau }_{\theta ,\min }}^{%
\widetilde{\tau }_{\theta ,\max }}\widetilde{I}(\tau _\theta
)(1+\beta \tau
_\theta )^2(1-\tau +\tau _\theta )d\tau _\theta \int_{\nu _1}^{\nu _2}\frac{%
g_{0,\nu }(\nu _{0,\theta })}\nu d\nu }{hcD^2\Gamma ^3(1-\beta
)^2(1+k\tau )^2},
\end{equation}
where $\widetilde{I}(\tau _\theta )$ represents the development of
the intensity magnitude in the observer frame, called as a local
pulse function,  $g_{0,\nu }(\nu _{0,\theta })$ describes the rest
frame radiation mechanisms, and $\tau$ is confined by $ 1-\cos
\theta _{\min }+(1-\beta \cos \theta _{\min })\tau _{\theta ,\min
}\leq \tau \leq 1-\cos \theta _{\max }+(1-\beta \cos \theta _{\max
})\tau _{\theta ,\max } $, while $\widetilde{\tau }_{\theta ,\min
}$ and $\widetilde{\tau }_{\theta ,\max }$ are determined by $
\widetilde{\tau }_{\theta ,\min }=\max \{\tau _{\theta ,\min
},\frac{\tau -1+\cos \theta _{\max }}{1-\beta \cos \theta _{\max
}}\} $ and $ \widetilde{\tau }_{\theta ,\max }=\min \{\tau
_{\theta ,\max },\frac{\tau -1+\cos \theta _{\min }}{1-\beta \cos
\theta _{\min }}\}, $, respectively. we will employ equation (1)
in the following analysis.
\subsection{In the case of radiation emitted from the whole fireball surface }
If the radiation is emitted from the whole fireball surface, we
can take  $\theta_{min}$ = 0  and  $\theta_{max} = \pi/2. $
Therefore the range of $\tau$ is $(1-\beta )\tau _{\theta
,\min}\leq \tau \leq 1+\tau _{\theta ,\max },$ while
$\widetilde{\tau }_{\theta , \min }$ and $\widetilde{\tau
}_{\theta , \max }$ are determined by $\widetilde{\tau }_{\theta ,
\min }=max\{\tau_{\theta , \min}, \tau-1\}$ and $\widetilde{\tau
}_{\theta , \max }=min\{\tau_{\theta , \max}, \tau/(1-\beta) \},$
respectively.

Formula (1) suggests that, except the state of the fireball (
i.e., $\Gamma$, $R_c$ and D), light curves of sources depend on
$\widetilde{I}(\tau _\theta )$ and $g_{0,\nu }(\nu _{0,\theta })$.
We assume in this paper the common empirical radiation form of
GRBs as the rest frame radiation form, the so-called Band function
(Band et al. 1993) that was frequently, and rather successfully
employed to fit the spectra of sources (see, e.g., Schaefer et al.
1994; Ford et al. 1995; Preece et al. 1998, 2000). We focus our
attention to the roles that the rising and decay portions of local
pulses play in the light curve of gamma-ray bursts. Several forms
of local pulses would be employed to study
this issue.\\
(a) Local pulse with a power law rise and a power law decay\\
A power law form of $\widetilde{I}(\tau _\theta )$ is adopted as
follows:
\begin{equation}
\widetilde{I}(\tau _\theta )=I_0\{
\begin{array}{c}
(\frac{\tau _\theta -\tau _{\theta ,\min }}{\tau _{\theta ,0}-\tau
_{\theta ,\min }})^\mu \qquad \qquad \qquad \qquad \qquad (\tau
_{\theta ,\min }\leq
\tau _\theta \leq \tau _{\theta ,0}) \\
(1-\frac{\tau _\theta -\tau _{\theta ,0}}{\tau _{\theta ,\max
}-\tau _{\theta ,0}})^\mu \qquad \qquad \qquad \qquad (\tau
_{\theta ,0}<\tau _\theta \leq \tau _{\theta ,\max })
\end{array}
\end{equation}
For this local pulse we find that $\Delta \tau _{\theta
,FWHM}=(1-2^{-1/\mu})(\tau_{\theta ,\max }-\tau_{\theta ,\min })$
and therefore, $\tau _{\theta ,max}=\frac{\Delta \tau
_{\theta,FWHM}}{(1-2^{-1/\mu})}+\tau_{\theta ,\min }$.

For a given value of $\tau _{\theta ,max}$, the rising timescale,
$\Delta\tau_{\theta,r}$, and the decay timescale
$\Delta\tau_{\theta,d}$ of a local pulse vary with the shift of
the position of $\tau _{\theta ,0}$, where $\Delta\tau_{\theta,r}$
and $\Delta\tau_{\theta,d}$ are the FWHM widthes in the rising and
decay portions, respectively. We assign
\begin{equation}
 \tau _{\theta,0}=\tau_{\theta ,\min }+e(\tau _{\theta ,max}-\tau _{\theta
 ,min}),  where  0 \leq e \leq 1.
\end{equation}
We find that $\Delta\tau_{\theta,r}=\tau _{\theta,0}-\tau
_{\theta,min}$, $\Delta\tau_{\theta,d}=\tau _{\theta,max}-\tau
_{\theta,0}$. Let
\begin{equation}
 r_{rd}(e)=\frac{\Delta\tau_{\theta,r}}{\Delta\tau_{\theta,d}}
\end{equation}
Quantity $r_{rd}(e)$ denotes the ratio of the rise timescale to
the decay timescale of a local pulse.

To describe different contributions of $\Delta\tau_{\theta,r}$ and
$\Delta\tau_{\theta,d}$ of a local pulse to the light curve, we
define
\begin{equation}
C_r(\tau )=\frac{2\pi R_c^3\int_{\widetilde{\tau }_{\theta ,\min }}^{%
\widetilde{\tau }_{\theta ,\max }}\widetilde{I}(\tau _\theta
)(1+\beta \tau
_\theta )^2(1-\tau +\tau _\theta )d\tau _\theta \int_{\nu _1}^{\nu _2}\frac{%
g_{0,\nu }(\nu _{0,\theta })}\nu d\nu }{hcD^2\Gamma ^3(1-\beta
)^2(1+k\tau )^2}, when   ( \tau _{\theta} \leq \tau _{\theta,0})
\end{equation}
\begin{equation}
C_d(\tau )=\frac{2\pi R_c^3\int_{\widetilde{\tau }_{\theta ,\min }}^{%
\widetilde{\tau }_{\theta ,\max }}\widetilde{I}(\tau _\theta
)(1+\beta \tau
_\theta )^2(1-\tau +\tau _\theta )d\tau _\theta \int_{\nu _1}^{\nu _2}\frac{%
g_{0,\nu }(\nu _{0,\theta })}\nu d\nu }{hcD^2\Gamma ^3(1-\beta
)^2(1+k\tau )^2}, when   ( \tau _{\theta} > \tau _{\theta,0})
\end{equation}
\begin{equation}
r_r =\frac{C_r(\tau )}{C(\tau )}
\end{equation}
\begin{equation}
r_d =\frac{C_d(\tau )}{C(\tau )}
\end{equation}
where $ C(\tau )=C_r(\tau )+C_d(\tau ).$ Quantities $C_r(\tau )$
and $C_d(\tau )$ denote the contributions of the rising timescale
and decay timescale of a local pulse to the light curve,
respectively.

Notes that when $\mu =1$, the local pulse (2) would become a
linear rising and a linear decay one.

Let us employ the Band function radiation form with $\alpha_0=-1$
and $\beta_0=-2.25$, to make the light curve, within the frequency
range of 100$\leq$ $\nu/\nu_{0,p}$$\leq$300. In order to
 obtain a more detailed information of a light curve profile, we evaluate
 the first order derivatives, $C'_r(\tau )$, $C'_d(\tau)$, $C'(\tau)$,
  and the second order derivatives $C''_r(\tau )$, $C''_d(\tau)$, $C''(\tau)$,
  of the count rate  $C_r(\tau )$, $C_d(\tau )$, $C(\tau)$,
   respectively. Their profiles are presented in Fig.1.

As shown in Fig. 1, two phases, A-B and D-E, undergo a concave
process as $C''_{AB}(\tau)$ $>$ 0 and $C''_{DE}(\tau)$ $>$ 0,
whereas other two phases, B-C and C-D, evolve a convex one because
of $C''_{BC}(\tau)$ $<$0 and $C''_{CD}(\tau)$ $<$0. We define
\begin{equation}
r_{cv} =\frac{\tau_{AB}}{\tau_{BC}}
\end{equation}
to denote the ratio of the two time intervals from concavity to
convexity in the rising portion of a light curve, where subscripts
``c" and ``v" represent concavity and convexity respectively.

We notice that $r_{cv}$ increases linearly with $r_{rd}$. For
example, $r_{cv}$ $\simeq$ 0.28 when $r_{rd}$=0.2, $r_{cv}$
$\simeq$ 0.76 when $r_{rd}$=0.5, and $r_{cv}$ $\simeq$ 1.64 when
$r_{rd}$=1. Similarly, $r_r$ increases, from $r_r < r_d$ to $r_r >
r_d$, with $r_{rd}$. We notice that there is an interesting point,
B, which is an inflexion from concavity to convexity in the rising
phase of the light curve, corresponding to the peak point in the
local pulse. Then the rising timescale $\Delta\tau_{\theta,r}$ of
a local pulse could be obtained by the location of point B from
the corresponding light curve.

We find that both the rising timescale and decay portions of the
local pulse make a contribution to both portions of the light
curve. The rising portion of the local pulse produces concave
phases on both sides of the light curve and creates a spiky
profile, whereas the decay one yields a convex phase in the rising
portion, and a convex-to-concave phase in the decay portion of the
light curve, and gives birth to a smooth profile. In a word, the
combinational contributions of both portions of the local pulse
lead to a light curve with its rising portion evolving from a
concave phase to a convex one (the so-called ``concave-to-convex"
character), and with its decay portion undergoing an opposite
process (the so-called ``convex-to-concave" one).\\
(b) Local pulse with a Gaussian form\\
Here, we replace $\widetilde{I}(\tau _\theta )$ in equation (1)
with a Gaussian form, which is as follows:
\begin{equation}
\widetilde{I}(\tau _\theta )=I_0\exp [-(\frac{\tau _\theta -\tau _{\theta ,0}%
}\sigma )^2]\qquad \qquad \qquad \qquad \qquad (\tau _{\theta
,\min }\leq \tau _\theta ).
\end{equation}
The profiles of the light curve are presented in Fig. 2.

We find that a local pulse with a Gaussian form also produces such
a profile of light curves: a ``concave-to-convex" shape in the
rising portion and a reverse one in the decay portion. The
differences are that a Gaussian form makes a smaller curvature
than a power law one does in the rising portion of light curves
either for the concavity or convexity phases, and that the value
of $r_{cv}$ keeps to be 1.24 because $r_{rd}$ keeps to be 1 for a
Gaussian form. However the values of $r_r$ and $r_d$ change with
$\sigma$.

(c)Local pulse with a $\delta$ function\\
Let us consider the case of $\widetilde{I}(\tau _\theta )$ being a
local $\delta$ function, which is as follows:
\begin{equation}
\widetilde{I}(\tau _\theta )=\frac{cI_0}{R_c}\delta (\tau _\theta
-\tau _{\theta ,0})\qquad (\tau _{\theta ,\min }\leq \tau _\theta
\leq \tau _{\theta ,\max })
\end{equation}

Associated with a local $\delta$ function pulse, the profile of
the light curve can be found in Fig. 1 of Qin et al. (2004). There
is no rising portion in the light curve for neither rising nor
decay timescale in the $\delta$ local pulse. But there is a decay
portion due to the curvature effect of the expanding fireball.

Besides the three local pulses discussed above, we also study
other forms of local pulses, such as $\mu$=2, 3 in equation (2),
as well as an exponential form of local pulses, and so on. The
same conclusions hold for all these case (see Fig. 1 and 2). The
value of $r_{cv}$ increases linearly with $r_{rd}$ at least in the
case of these local forms we investigate. The curvature of the
rising portion also changes with different forms of local pulses
and different $r_{rd}$, which are responsible for the diversity of
light curves.
\subsection{In the case of radiation emitted from a uniform jet  }
If the radiation is emitted from a jet-like geometry with a
half-opening angle $\theta_{j}$ and a spherical surface (a uniform
jet), we assume for such a relativistic outflow both the Lorentz
factor $\Gamma$ and the energy per unit solid angle from the jet
axis are constant during the GRB phase, and assume that the
observer direction is the jet axis. At these assumptions, we are
able to apply equation (1) to the uniform jet taking
$\theta_{min}=0$ and $\theta_{max}=\theta$, where $\theta$
represents the half-opening angle of the jet.

With $\Gamma$=10 and 100, $\theta_{min}=0$ and
$\theta_{max}=1/\Gamma$, we repeat the same work as does in the
$\S$ 2.1, and find the same characteristics of a light curve of
GRBs: a ``concave-to-convex" phase in the rising portion and a
reverse phase in the decay portion, except that there exist a
cutoff tail feature in the decay portion (see Fig.1 in Qin et al.
2004). (the cutoff tail feature does not appear in a light curve
when $\theta_{max}$ is large enough). The characteristics appear
to be the result of the curvature effect regardless of a spherical
fireball or a uniform jet as long as its surface, from which the
radiation is emitted, is a spherical one.

\section{Observational Data}
To test our analysis mentioned above, we examine the profiles of
light curves of GRBs detected by the BATSE instrument on board the
CGRO (Compton Gamma Ray Observatory). It is already known that
pulses of a GRB show a tendency to self-similarity for different
energy band (see, e.g. Norris et al. 1996). Therefore, we study in
this paper only the count rate of channel 3.

For each source, the background is estimated by a fit with a
polynomial expression using 1.024s resolution data that are
available from 10 minutes before the trigger to several minutes
after the burst. The data along with the background fit
coefficients can be obtained from the CGRO Science Support Center
(CGROSSC) at NASA Goddard Space Flight Center through its public
archives. After subtracting the background counts, we smooth the
data with the DB3 wavelet with the MATLAB software in the level of
the third-class decomposition. We select a sample comprising of 86
well-separated pulses from these data, on which almost no other
small pulses overlap.

For the sake of comparison, the selected pulses above are
normalized and their variables, t, are re-scaled so that the peak
count rate is located at t=0 and the FWHM point of the decay
portion is located at t=0.2. To obtain the value of $r_{cv}$ in a
pulse, we evaluate the first order derivatives, $C'(\tau)$, of a
pulse, and then identify the inflexion from concavity to
convexity. Applied this method, we receive a list of the value of
$r_{cv}$ for our sample pulses, which is shown in Table 1. The
distribution of $r_{cv}$ is presented in Fig. 2.
\begin{table*}
\centering \caption{A list of the value of $r_{cv}$ for our sample
pulses}
\begin{tabular}{|c|c||c|c||c|c|}
\hline\hline trigger number & $r_{cv}$ &trigger number & $r_{cv}$ &trigger number & $r_{cv}$\\
\hline
       493 &    0.70 &       2665 &    0.80 &       5474 &    1.15 \\
       563 &    0.91 &       2700 &    3.53 &       5478 &    0.50 \\
       829 &    0.96 &       2749 &   11.29 &       5495 &    0.93 \\
       907 &    0.69 &       2880 &    7.85 &       5517 &    1.70 \\
       914 &    2.05 &       2919 &    1.48 &       5523 &    7.47 \\
       973 &    0.99 &       3040 &    1.31 &       5601 &    1.02 \\
       973 &    0.80 &       3143 &    6.19 &       5601 &    1.52 \\
       999 &    4.17 &       3155 &    0.76 &       6225 &    0.31 \\
      1157 &    7.75 &       3256 &    1.63 &       6335 &    2.84 \\
      1406 &    0.69 &       3237 &    1.44 &       6397 &    1.02 \\
      1467 &    2.71 &       3257 &    1.23 &       6504 &    0.70 \\
      1700 &    1.11 &       3415 &    0.66 &       6621 &    2.00 \\
      1733 &    5.01 &       3648 &    3.58 &       6672 &    1.26 \\
      1883 &    1.69 &       3668 &    2.87 &       6672 &    0.82 \\
      1956 &    1.56 &       3765 &    1.92 &       6930 &    1.70 \\
      1989 &    0.11 &       3866 &    4.20 &       7293 &    1.46 \\
      2083 &    4.68 &       3870 &    1.29 &       7295 &    1.51 \\
      2102 &    2.41 &       3875 &    0.63 &       7374 &    0.42 \\
      2138 &    1.99 &       3886 &    4.50 &       7403 &    0.65 \\
      2267 &    4.52 &       3892 &    0.85 &       7548 &    1.18 \\
      2306 &    4.90 &       3954 &    7.85 &       7548 &    0.70 \\
      2387 &    0.70 &       4157 &    1.36 &       7588 &    0.90 \\
      2393 &    7.28 &       4350 &    0.90 &       7638 &    3.80 \\
      2393 &    2.99 &       4350 &    2.75 &       7648 &    1.45 \\
      2431 &     6.0 &       4350 &    2.52 &       7701 &    4.72 \\
      2484 &    1.12 &       4368 &    1.02 &       7711 &    0.38 \\
      2519 &    4.07 &       4368 &    0.74 &       7744 &   10.74 \\
      2530 &    0.37 &       4710 &    0.89 &       8111 &    2.40 \\
      2662 &    1.19 &       5415 &    1.39 &            &         \\
 \hline
\end{tabular}
\end{table*}

\begin{figure}
  \includegraphics[width=5in,angle=0]{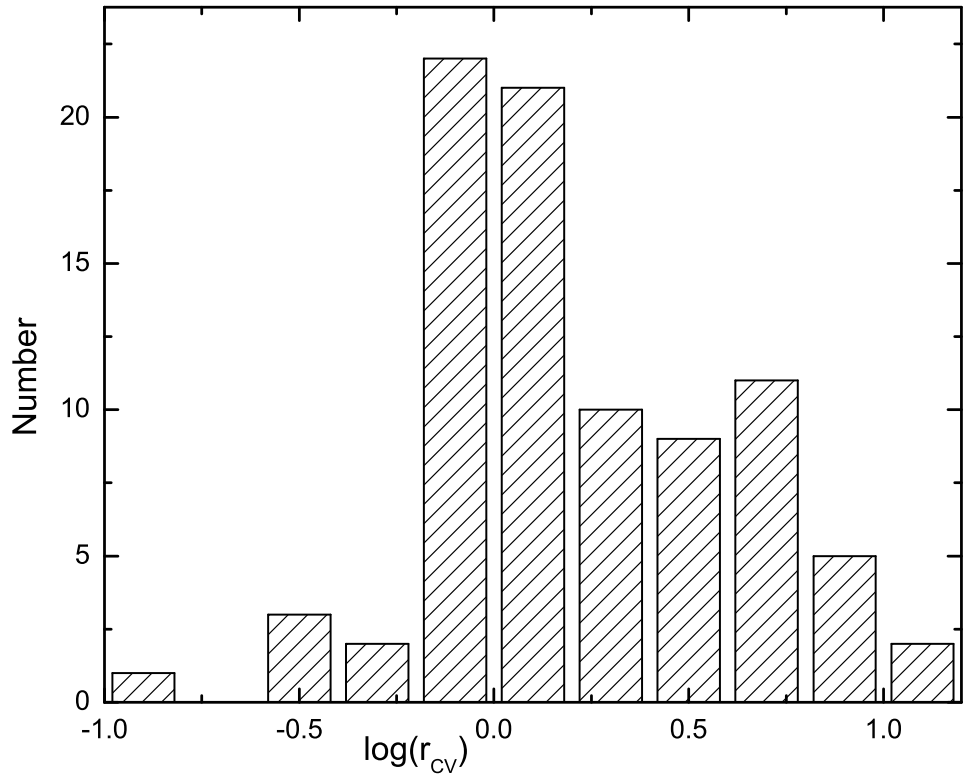}
  \caption{Distribution of $r_{cv}$ of all pulses in our sample}
 \label{fig3}
 \end{figure}

\begin{figure}
  \includegraphics[width=5in,angle=0]{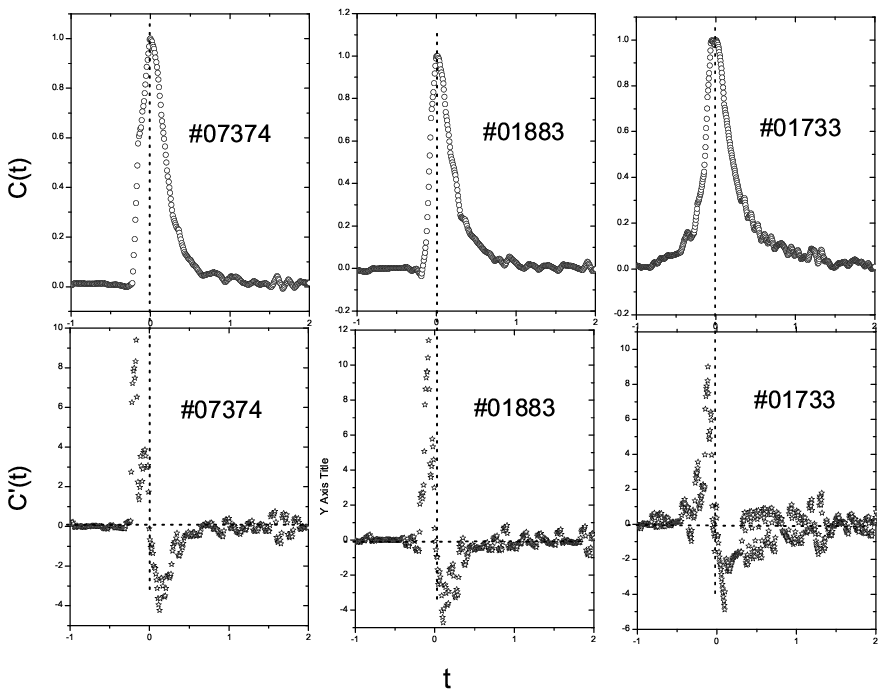}
  \caption{Plot of the normalized and re-scaled pulses
  (first row) and their slopes (second row) for several sample source. }
 \label{fig4}
 \end{figure}
According to Fig. 3, we divide our sample into 3 classes by the
range of $r_{cv}$, which are named as class 1 ($r_{cv}$ $\leq$
0.5), class 2 (0.5 $<$ $r_{cv}$ $\leq$ 2) and class 3 ($r_{cv}$
$>$ 2). Their percentages to the total number are 5.8, 58.1 and
36.1, respectively. The observed distribution of $r_{cv}$, whose
explanation is not clear, may constrain the mechanism of the birth
of local pulses in both models.

Light curve and their slopes of three events, $\sharp$ 07374
($r_{cv}=0.42$), $\sharp$ 01883 ($r_{cv}=1.69$) and $\sharp$ 01733
($r_{cv}=5.01$), are drawn in Fig. 4. One find that the characters
of the slopes are in agreement with what we obtain in last
section. This suggests that, at least for our sample sources, the
curvature effect (or the Doppler effect of an expanding spherical
surface) for a uniform jet or a spherical fireball is indeed at
work.
\section{discussion and conclusions  }
All analyses in this paper are based on formula (1), which is
suitable for describing light curves of spherical fireballs or
uniform jets. Numerical calculations show that the characters of
light curves, the ``concave-to-convex" one in the rising phase and
the ``convex-to-concave" one in the decay portion, are not
affected by the Lorentz factor $\Gamma$ and the half-opening angle
$\theta_{j}$.

Let us study the impact of rest frame radiation forms on a light
curve profile. As a general radiation form observed, the Band
function, for which, some sets of typical values of the indexes
would represent certain mechanisms (see Band et al. 1993), will be
employed. Here we take three sets of values of the indexes:
$\alpha_0$=-1 and $\beta_0$=-2.25, $\alpha_0$=-1 and
$\beta_0$=-4.5, $\alpha_0$=-0.5 and $\beta_0$=-2.25. The light
curves, slopes, $r_r$ and $r_d$, are plotted in Fig.5.
\begin{figure}
\includegraphics[width=5in,angle=0]{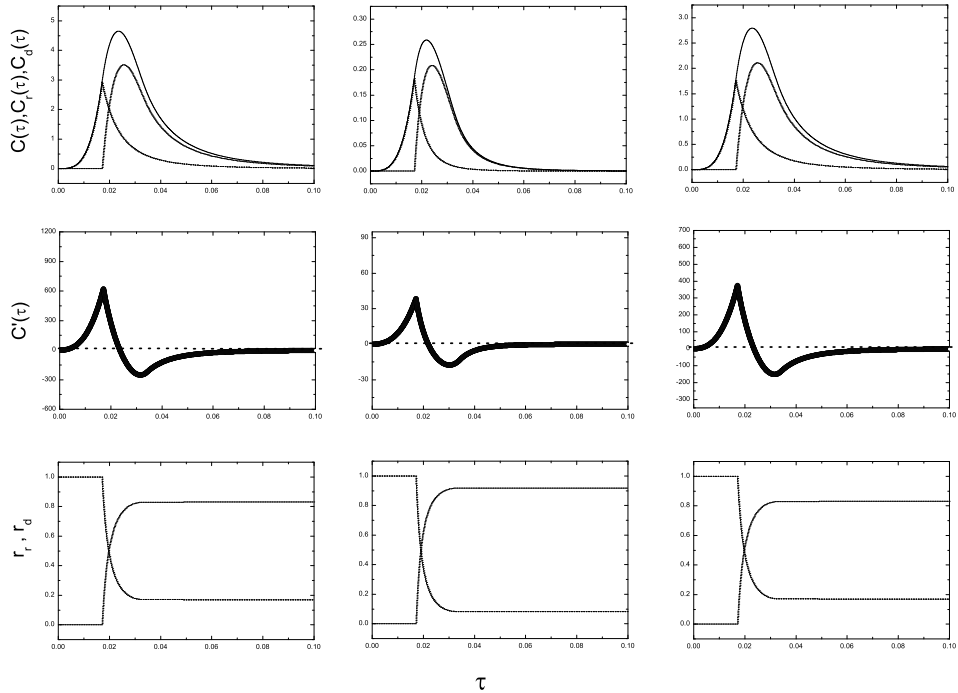}
\caption{Plot of $C(\tau)$, $C_{r}(\tau)$, $C_{d}(\tau)$---$\tau$,
$C'(\tau)$---$\tau$, $r_r$, $r_d$--$\tau$ for the light curve of
the power law local pulse determined by (2), where a Band function
rest frame form is adopted. And we take $\Gamma$=10, $\mu=2$,
$\Delta\tau_{\theta,FWHM}$=2, $\tau_{\theta,min}$=0,
$\tau_{\theta,max}$=2. $\tau_{\theta,0}$=$\tau_{\theta,max}$*0.5.
$\alpha_0$=-1 and $\beta_0$=-2.25 ( for first column ),
$\alpha_0$=-1  and $\beta_0$=-4.5 (for second column),
$\alpha_0$=-0.5 and $\beta_0$=-2.25 (for last column). The symbols
are the same as those adopted in Fig.1.}
 \label{fig5}
 \end{figure}

As shown in the figure, different rest frame radiation forms lead
to different values of $r_r$ and $r_d$, but the value of $r_{cv}$
and the profile character mentioned above are not affected.

Zhang et al. (2002) found that a structured jet has the same
temporal evolutions as the uniform jet has as long as the
relativistic beaming angle 1/$\Gamma$ is much smaller than the
viewing angle $\theta_{v}$, in this situation, the fireball is
observed as if it were isotropic. We thus suspect that the
characteristics of light curves in the structure jet may probably
be similar to those obtained above, which deserves a detailed
analysis.

One may ask whether or not the characteristics of the profile of
light curves come from the intrinsic behavior. Our analysis shows
that it is not the case. Taking $\mu=1$ in equation (2), which
become a linear rise and a linear local pulse, the characters are
the same as in the case of $\mu=2$. The fact shows that the
characteristics indeed result from the curvature effect, instead
of being an intrinsic one.

We come to the conclusion that, the light curve characteristics, a
``concave-to-convex" curve in the rising portion and a revers
curve in the decay portion, which is independent of the shapes of
local pulses, exist in the observed light curve of some GRBs (We
find 86 source in our sample). We believe that reverse characters
in both portions (Say, a ``convex-to-concave" curve in the rising
portion and a ``concave-to-convex" curve in the decay part) would
never be detected if the radiated surface is a spherical one,
regardless of a spherical fireball or a uniform jet. The
characters could serve as an indicator of the curvature effect or
the Doppler effect.

This work was supported by the Special Funds for
Major State Basic Research Projects (``973'') and National Natural
Science Foundation of China (No. 10273019).

\label{lastpage}

\end{document}